\documentclass[prb]{revtex4}
\usepackage{amsmath}
\usepackage{bookmark}
\newcommand{\be}{\begin{equation}}
\newcommand{\ee}{\end{equation}}
\begin{document}

\title{Fundamental invariants of many-body Hilbert space}

\author{D. K. Sunko}

\affiliation{Department of Physics,
Faculty of Science, University of Zagreb, Bijeni\v{c}ka 32,\\
10000 Zagreb, Croatia}

\email{dks@phy.hr}

\begin{abstract}

Many-body Hilbert space is a functional vector space with the natural
structure of an algebra, in which vector multiplication is ordinary
multiplication of wave functions. This algebra is finite-dimensional, with
exactly $N!^{d-1}$ generators for $N$ identical particles, bosons or fermions,
in $d$ dimensions. The generators are called \emph{shapes}. Each shape is a
possible many-body vacuum. Shapes are natural generalizations of the
ground-state Slater determinant to more than one dimension. Physical states,
including the ground state, are superpositions of shapes with
symmetric-function coefficients, for both bosons and fermions. These symmetric
functions may be interpreted as bosonic excitations of the shapes. The
algebraic structure of Hilbert space described here provides qualitative
insights into long-standing issues of many-body physics, including the fermion
sign problem and the microscopic origin of bands in the spectra of finite
systems.

\end{abstract}

\keywords{Many-body wave functions; fermion sign problem; invariant theory.}

\maketitle
\section{Introduction}

The problem of constructing many-body wave functions of identical particles is
almost as old as quantum mechanics itself. Soon after Pauli\cite{Pauli25}
articulated his famous principle, Heisenberg\cite{Heisenberg26} and
Dirac\cite{Dirac26} found that it could be mathematically expressed by wave
functions in determinantal form, which came to be known as
Slater\cite{Slater29} determinants. Given that Slater determinants form a
complete basis of Hilbert space, the so-called second-quantized formalism,
which made their manipulation technically feasible even for large systems,
established them as a secure formal foundation for all microscopic
descriptions of many-fermion systems.

While formally universal, the Slater-determinant approach --- as implemented
most characteristically in the time-dependent pertubation
formalism\cite{Abrikosov75,Mahan90} --- has a pattern of success indicating
that it is better adapted to some problems than others. Broadly speaking, it
works when the interacting system can be described in terms of relatively
simple excitations of the non-interacting system. Even an infinity of such
excitations, creating collective modes and new ground states, is within the
purview of the approach, provided that each individual excitation can be
expressed as a small departure from the non-interacting ground state. The
paradigm for such small departures are the well-known particle-hole
excitations of the Fermi sea. Similarly, Cooper pairing is just a resonance in
two-particle head-on collisions.\cite{Mahan90}

On the other hand there are so-called strongly correlated systems, for which
the Slater determinant formulation does not seem to offer advantages by
itself. These are characterized by a number of properties which need not all
be present, but typically include:
\begin{itemize}
 \item a finite number of particles;
 \item localization in space;
 \item strong short-range or unscreened long-range interactions.
\end{itemize}
All three characteristics are present in atoms, nuclei, and small molecules.
The two-dimensional electron gas, found in the inversion layers of
semiconducting heterostructures, has only the third property in the absence of
a magnetic field, but acquires the other two when a strong field is turned on.
Of course, the number of electrons does not suddenly become finite, rather the
localization of trajectories caused by the field means that only a small
number of electrons create a strongly correlated wave function, which is then
repeated throughout space. This number was three in Laughlin's original
description\cite{Laughlin83-1} of the fractional quantum Hall effect (FQHE) at
$1/3$ filling, where the strongly correlated motion in question was their
rotation as the vertices of a ``stiff'' equilateral triangle. Some
contemporary approaches\cite{Hyowon14} to strongly correlated materials share
the same pattern: the local part of the problem is treated by dynamical
mean-field theory\cite{Georges96} (DMFT), while the extension in space is
treated by density-functional theory\cite{Kohn65} (DFT), i.e., by Slater
determinants. A significant difference between the latter two cases is that
all electrons need to be antisymmetrized in the FQHE, while in crystals the
electrons localized within a unit cell need not be antisymmetrized with those
in the other cells. The separation between the local and the itinerant degrees
of freedom in such cases is an important issue beyond the scope of the present
review.

Here, the first tentative steps of a new approach are described, which shows
some promise to be ``native'' for few-particle strongly correlated cases, much
as the Slater determinants natively describe long-wavelength excitations of
extended systems. The present review places this approach in its natural
mathematical context and addresses various physical ramifications in a
qualitative way. Proofs of the main results have appeared
elsewhere.\cite{Sunko16-1} The primary ambition, unfulfilled at present, is to
obtain \emph{realistic} wave functions of few-body systems, without
necessarily improving on existing numerical methods either in terms of speed
or accuracy in the calculation of binding energies. In this sense the
philosophy is converse to that of DFT, which consciously sacrifices realism of
the wave functions in favor of precise calculation of energies.

\section{Mathematical background}

The mathematics invoked in the present work has a long and distinguished
lineage. Combinatorics and the theory of invariants were first brought into
modern form by Leonhard Euler and David Hilbert, respectively. The purpose of
the present section is only to introduce as much background as necessary to
understand the subsequent physical development. The level is chosen to make the
material accessible to any physics student. If the mathematical training of
physicists were devoted half as much to abstract algebra as it is to complex
analysis, the whole section would be unnecessary. The style is informal, with
mathematical rigor left to the references. Although every statement is true as
it stands, the presentation is tendentious in light of the final purpose ---
describing the physical Hilbert space of $N$ particles --- so that it should not
be taken to reflect the attitudes of mathematicians.

\subsection{Euler and generating functions}

Euler invented generating functions\cite{Euler51} in 1741, ostensibly to
answer a question from a ``most insightful'' correspondent: in how many ways
can the number $50$ be written as a sum of $7$ distinct natural numbers? Of
course, permutations of the same sum are not counted, so the question may be
rephrased, ``\ldots sum of $7$ \emph{strictly increasing} natural numbers.''
Euler realized that if he could compute the sum
\begin{equation}
\sum_{1\le n_1<n_2<n_3<n_4<n_5<n_6<n_7<\infty}
q^{n_1+n_2+n_3+n_4+n_5+n_6+n_7},
\label{original}
\end{equation}
the solution would appear as the coefficient of $q^{50}$. In terms of
contemporary physics, his task was to compute the degeneracy of the level with
energy $50\,\hbar\omega$ of a one-dimensional harmonic oscillator well, filled
with $7$ fermions, and spectrum given by $\varepsilon_n=n\hbar\omega$, $n\ge
1$, and $q=\exp(-\beta\hbar\omega)$. (The answer is $522$.)

The sum~\eqref{original} is simple enough to be computed directly. By some
clever tricks with indices\cite{Andrews76} one can find that\cite{Euler51}
\be
\sum_{0\le n_1<\ldots<n_N<\infty}q^{n_1+\ldots +n_N}
=q^{N(N-1)/2}\frac{1}{1-q}\frac{1}{1-q^2}\cdots\frac{1}{1-q^N}
\equiv q^{N(N-1)/2}Z_E,
\label{Z1}
\ee
where the origin of the $n_i$ is shifted to zero for later convenience (the
original sum yields the prefactor $q^{N(N+1)/2}$). Interestingly, if one
changes the restrictions on the indices to $\le$ instead of $<$, which means
replacing fermions with bosons, only the prefactor changes to reflect the
different boson ground-state energy. Fermions and bosons are not really
different in one dimension, an observation whose proper generalization to
$d>1$ will become apparent below.

Euler did not compute the sum directly. Instead he presented the result
\eqref{Z1} constructively, as just one particular manipulation of the simpler
generating functions, appearing as terms in the product. He inaugurated the
theory of partitions by extending such manipulations to prove his famous
``pentagonal number theorem,'' a development\cite{Euler51,Andrews76} which
need not concern us here. We shall however need another of his observations:
upon expanding and multiplying the geometric series in the product
\begin{equation}
\frac{1}{1-e_1}\frac{1}{1-e_2}\cdots\frac{1}{1-e_N},
\label{eulergen}
\end{equation}
each distinct monomial $e_1^{n_1}e_2^{n_2}\cdots e_N^{n_N}$ appears exactly
once. While the observation itself may appear obvious enough, it marks a
historical shift towards an abstract use of mathematical symbols, in which
``letters'' are not considered as place-holders for numbers, but as
mathematical objects in their own right. The observation makes sense to one
who is prepared to use a mathematical expression like \eqref{eulergen} as a
sort of symbolic machine to generate other expressions, namely the said
monomials. The statement~\eqref{eulergen} is then a precise prescription to
manipulate symbols, not any numerical value. Here it will be used as a working
part of a bigger symbolic machine, to generate expressions for the shape wave
functions.

\subsection{Hilbert and invariants}

While a particular date may be assigned to the invention of generating
functions, invariants have been present in mathematical thought since the
earliest times. As soon as they were discovered in the 4th century BC, conic
sections were classified according to how the plane cutting the cone was
tilted: too little (ellipse), just enough (parabola), or too much (hyperbola).
The invariant nature of this classification is intuitively obvious to anyone
looking at the intersection from the outside. A significant step forward was
to realize that a small bug living in the plane could also distinguish the
three cases, by fitting the curve to the equation $Ax^2+Bxy+Cy^2+Dx+Ey+F=0$
and computing the discriminant $B^2-4AC$. This program can be carried out in
two ways. One is to study a small neighborhood of a point on a curve, leading
eventually to differential geometry. The discriminant is then just the Hessian
of the quadratic form, describing the local curvature of the conic.

An alternative to the differential-geometric is the algebraic point of view,
which notes that the fitting needs to evaluate the curve only at a finite
number of points. The coefficients are now obtained by an investigation
limited in a different sense: instead of being continuous in an infinitesimal
neighborhood, it is discrete in a finitely extended neighborhood. The
polynomial is nothing but its list of coefficients, which are numbers in a
certain relationship. The relationship is best presented by keeping the old
notation for a polynomial, but now the variables $x$ and $y$ are just letters,
or notational tags, to keep track of the roles played by the coefficients,
e.g. when calculating the discriminant. Relabeling the tags by a Euclidean
transformation of the pair $(x,y)$ changes the coefficients without changing
the discriminant: it is said to be an \emph{invariant} of the quadratic form
under action of the Euclidean group in the plane. In this ``classical''
program, usually associated with Cayley, the main concern was how to find all
the invariants of a particular group action in any given case.

If Euler had opened the study of generating functions, Hilbert was widely
credited with closing the classical study of invariants, by proving a
structural result which relegated the computation of particular invariants to
applications. His attitude can be introduced in a succession of steps. First,
there is a mapping between the coefficients and roots of a polynomial, namely
if the roots are $t_i$, $i=1,\ldots N$, the coefficients are $(-1)^ke_k$,
where $e_k$ are the elementary symmetric functions,
\begin{align}
e_1&=t_1+\ldots+t_N=\sum_it_i,\nonumber\\
e_2&=t_1t_2+t_1t_3+\ldots+t_{N-1}t_N=\sum_{i<j}t_it_j,\nonumber\\
&\vdots\nonumber\\
e_N&=t_1t_2\cdots t_N.\label{ek}
\end{align}
These are just the Vi\`ete formulas, first encountered in high-school algebra.
All functions of the coefficients, including the invariants such as the
discriminant, are thus necessarily symmetric functions of the roots. Taking
the roots as the new variables, one parametrizes all the polynomials at once.
Instead of the old dummy variables $x,y$ being used as handles for group
action in the geometric ``outer'' space over which the polynomials were
defined, attention shifts to the abstract ``inner'' space of the polynomials
themselves, as parametrized by their roots. This crucial change of focus is
almost imperceptible in the notation, because of course the roots may be
visualized as points in the very same geometric space in which the original
polynomial was being evaluated. In the mathematical literature, it is not
uncommon to treat the $e_i$ themselves as free parameters, given that they
parametrize the polynomials equally well. As will appear below, in physics the
difference between the two must remain marked, because roots are related to
coefficients like one-body to many-body wave functions.

Taking the discriminant of a quadratic $x^2-e_1x+e_2$ as an example,
\begin{equation}
e_1^2-4e_2=(t_1-t_2)^2,
\end{equation}
one sees immediately that it is an invariant of the Euclidean group, which by
definition preserves distances. In fact the discriminant may in general be
defined as the square of the Vandermonde determinant\cite{Aitken39} $\Delta$,
\begin{equation}
\Delta(t_1,\ldots,t_N)\equiv
\left|\begin{matrix}
t_1^{N-1}&\cdots&t_N^{N-1}\\
\vdots&&\vdots\\
t_1&\cdots&t_N\\
1&\cdots&1
\end{matrix}\right|=
\prod_{1\le i<j\le N} (t_i-t_j).
\label{vandermonde}
\end{equation}
One immediate consequence of this point of view is that one has generated a
great many rather obvious invariants: because all functions of the
coefficients are symmetric in the roots, they are all invariant under the full
group of permutations of the $N$ roots, the symmetric group $S_N$. These may
be understood as invariants with respect to renaming the roots. They can all
be generated by Euler's expression~\eqref{eulergen}. To be explicit, the
monomials generated by~\eqref{eulergen} are a basis in the space of
invariants: any invariant of the symmetric group can be expressed as a linear
combination of these monomials, i.e.\ a polynomial in the $e_i$,
$i=1,\ldots,N$.

In the next step, obtaining ``interesting'' invariants becomes a program of
reduction of the complete monomial basis obtained above. This program cannot
miss any invariants, because \emph{any} finite group operating on the $N$
points $t_i$ can be understood as a subgroup of the permutation group, a
result due to Cayley. For example, some invariants under root reflection
$t_i\to-t_i$ are generated from the coefficients of even index $e_{2i}$, and
others from squares of coefficients of odd index, $e_{2i+1}^2$. The
corresponding reduced generating expression~\eqref{eulergen} is obtained
simply by squaring all the $e_i$ with odd indices appearing in the
denominators. However, this expression misses ``mixed'' reflection-symmetric
terms, such as $e_1e_3$. A moment's thought suffices to realize that the
generating function of all invariants of (say) cubic polynomials
$x^3-e_1x^2+e_2x-e_3$ under root reflection is just
\begin{equation}
\frac{1}{2}\left(\frac{1}{1-e_1}\frac{1}{1-e_2}\frac{1}{1-e_3}+
\frac{1}{1+e_1}\frac{1}{1-e_2}\frac{1}{1+e_3}\right)
=\frac{1+e_1e_3}{(1-e_1^2)(1-e_2)(1-e_3^2)},
\label{e123}
\end{equation}
because it lists all and only monomials which are unchanged by the change of
sign. In other words, any invariant under root reflection takes the form
\begin{equation}
R(e_1,e_2,e_3)=P(e_1^2,e_2,e_3^2)\cdot 1+Q(e_1^2,e_2,e_3^2)\cdot e_1e_3,
\label{PQ}
\end{equation}
where $P$ and $Q$ are arbitrary polynomials in three variables.

An expression like~\eqref{e123} or~\eqref{PQ} is the most general answer one
can give to a problem of finding invariants. Hilbert's conclusive contribution
was to realize not only that such expressions were always possible, but in
addition that the numerator in Eq.~\eqref{e123}, or the sum in Eq.~\eqref{PQ},
will be \emph{finite}, i.e.\ not an infinite series, for a very large class of
symmetry groups,\cite{Derksen02} so large that physicists need not be
concerned with the exceptions.

\subsection{The algebra of invariants}

The structure of Hilbert's insight is visible in Eq.~\eqref{PQ}: in some
sense, there are only two interesting invariants in the problem, namely the
set $\{1,e_1e_3\}$, listed in the numerator of Eq.~\eqref{e123}. By contrast,
$P$ and $Q$ are arbitrary, once one has identified their allowed variables.
This separation follows the natural reasoning above, which quickly identified
$e_{2i}$ and $e_{2i+1}^2$ as obvious invariants under root reflection, but was
hesitant how to include $e_1e_3$ properly. Terminologically, the monomials
appearing in the geometric series (or in $P$ and $Q$) are called
\emph{primary}, while those appearing in the polynomial in the numerator of
Eq.~\eqref{e123} are called \emph{secondary} invariants.\cite{Derksen02}

The form of the expression~\eqref{PQ} indicates that the secondary invariants
act as a basis, spanning the whole space of invariants. This insight is
formalized by noting that polynomials are a vector space over the real (or
complex) numbers, so that ordinary multiplication of polynomials is vector
multiplication in that space, therefore polynomials naturally form an algebra.
The basis vectors intuited above are just the generators of that algebra. Any
invariant $p$ may be written
\begin{equation}
p=\sum_{i=1}^Dh_ig_i,
\label{palg}
\end{equation}
where the $h_i$ are arbitrary polynomials in the primary invariants and $g_i$
are secondary invariants, or generators of the algebra of invariants. The sum
is always finite, by Hilbert, for all classical and finite groups acting on
polynomials over the complex field.\cite{Derksen02} It is said that such an
algebra is \emph{finitely generated}, where $D$ is its dimension [number of
its generators, $D=2$ in the example~\eqref{PQ}].

An additional structure can be inferred from Eq.~\eqref{palg}. Let $g_i'$,
$i=1,\ldots,M$, be some linear combinations of the generators $g_i$. Then
polynomials $q$ of the form
\begin{equation}
q=\sum_{i=1}^Mh_ig_i',
\label{ideal}
\end{equation}
with $h_i$ the same as before, form an \emph{ideal} in the space of invariants
$p$ in Eq.~\eqref{palg}. This proposition is easily checked from the
definition. Ideals are subsets closed under addition,
$q+q'=\sum(h_i+h_i')g_i'$, and absorptive under multiplication,
$h'q=\sum(h'h_i)g_i'$. An important question now is whether the $q$'s form a
proper subset of the $p$'s. Even if $M<D$, one cannot be sure that the
generators $g_i$ were really the minimal set needed to span the space of
$p$'s, because a clever choice of the $g_i'$ could in principle transfer some
of the burden to the coefficients $h_i$. In other words, because of the
flexibility of polynomial multiplication, the dimension of the algebra of
invariants can be difficult to divine in practice, leading to a large body of
research devoted to basis-reduction algorithms.\cite{Derksen02} Sometimes,
however, one can argue \emph{a fortiori} that the most efficient
representation has been achieved. Such is our running example~\eqref{PQ},
noting that the term $e_1e_3$ is even under root reflection but cannot appear
in $P$ or $Q$.

\subsection{Grading and Hilbert series}

Instead of explicitly generating all polynomial invariants, one may ask how
many of a given degree exist. Clearly each elementary symmetric function $e_k$
is of degree $k$. Hence the replacement $e_k\to q^k$ in the
example~\eqref{e123} gives the answer: the number of independent invariants of
degree $m$ is the coefficient of $q^m$ in the expression
\begin{equation}
\frac{1+q^4}{(1-q^2)^2(1-q^6)}.
\label{hilbex}
\end{equation}
Such a series is called a Hilbert (sometimes\cite{Stanley78} Poincar\'e)
series. The classification of polynomials by degree is a particular example of
a \emph{grading}, which is any decomposition of an algebra (or ring) in a
direct sum of subspaces $M_i$, labeled by natural numbers $i$, which has the
property $M_iM_j\subseteq M_{i+j}$.

The Hilbert series is an important practical tool for finding invariants,
because computing it is in principle much easier than finding a closed
invariant-generating function like Eq.~\eqref{e123}. Sometimes it gives
sufficient hints of the structure of the space of invariants to enable the
design of concrete algorithms for their construction. As we shall see now,
that is the case with many-body Hilbert space. 

\section{Main results}

Having established the mathematical language, all the main results concerning
many-body Hilbert space can be stated transparently. The proofs appear
elsewhere.\cite{Sunko16-1}

\subsection{Sum over states}

The natural grading for wave functions is the number of one-body nodes. In
order to make the node-numbers independent in the various coordinates, only
bases separable in Cartesian coordinates are considered. Then the most
efficient way to represent one-body wave functions is as pure powers of some
formal variables, with the exponent denoting the grade, e.g., $t^lu^mv^n$ in
three dimensions. In one dimension, the ground-state Slater determinant then
becomes just the Vandermonde determinant~\eqref{vandermonde}. Any $N$-body
wave function can be readily recovered from this representation:
\begin{equation}
t_i^lu_j^mv_k^n\to H_l(x_i)e^{-x_i^2/2}
H_m(y_j)e^{-y_j^2/2}H_n(z_k)e^{-z_k^2/2}
\end{equation}
for the 3D harmonic oscillator, say, with $i,j,k=1,\ldots,N$. In this mapping,
the one-body functions must not be individually normalized. Only the final
superpositions are normalized before concrete calculations.

The Hilbert series for this grading is just the physical partition function of
the harmonic oscillator, because the energy is linear in the number of nodes
in that case. Thus all oscillator counting results become completely general.
The price to pay is, of course, that wave functions classified by grade are
not necessarily the eigenfunctions of any Hamiltonian, as soon as one invokes
some other one-body realization than the oscillator one, say $\cos mx$ in
place of $H_m(x)e^{-x^2/2}$ above. However, that is a very small price,
because one does not hope anyway that the initial choice of basis will solve
any non-trivial problem. Replacing $e^{-\beta\hbar\omega}$ by the letter $q$,
the Hilbert series for $N$ identical particles in $d$ dimensions
reads\cite{Sunko16-1}
\begin{equation}
Z_d(N,q)=Z_E(N,q)^dP_d(N,q),\quad Z_E=\prod_{k=1}^N\frac{1}{1-q^k},
\label{zdnq}
\end{equation}
where $P_d(N,q)$ is a polynomial in $q$ which satisfies \emph{Svrtan's
recursion}
\begin{equation}
NP_d(N,q)=\sum_{k=1}^{N}(\pm 1)^{k+1}\left[C^{N}_{k}(q)\right]^dP_d(N-k,q),
\label{recurPd}
\end{equation}
with the upper sign for bosons, and the lower for fermions. Here
\be
C^N_k(q)=\frac{(1-q^N)\cdots (1-q^{N-k+1})}{(1-q^k)}
\ee
is a polynomial, and $P_d(0,q)=P_d(1,q)=1$.

The Hilbert series~\eqref{zdnq} already reveals a lot about the structure of
$N$-body Hilbert space. In particular, the polynomial $P_d(N,q)$ counts the
$d$-dimensional states which \emph{cannot} be induced from the one-dimensional
ones by simple combinatorial tricks, such as
\be
\Psi_a(x_1,\ldots,x_N)
\to\Psi_a(x_1,\ldots,x_N)\Psi_b(y_1,\ldots,y_N)\Psi_c(z_1,\ldots,z_N).
\label{triv}
\ee
Namely, these trivial extensions of 1D states are counted by the other factor,
$Z_E^d$. Furthermore, comparing Eqs.~\eqref{e123} and~\eqref{zdnq} shows that
$P_d(N,q)$ is the numerator in these expressions, hence it counts the
secondary invariants. Thus the form of Eq.~\eqref{zdnq} identifies $N$-body
Hilbert space as a finitely generated graded algebra. Because the number of
generators of each grade [coefficient of a power of $q$ in $P_d(N,q)$] must be
a nonnegative integer, inserting $q=1$ into the recursion~\eqref{recurPd}
gives the total dimension of the algebra, $P_d(N,1)=N!^{d-1}$. For example,
in the case of three-fermion wave functions in two dimensions,
$P_2(3,q)=1+4q+q^2$ and $P_2(3,1)=6=3!^1$.

\subsection{Wave functions}

The generators of the Hilbert-space algebra are called \emph{shapes}. Any
state in Hilbert space has a wave function of the form
\be
\Psi=\sum_{i=1}^{N!^{d-1}} \Phi_i\Psi_i,
\label{genspan}
\ee
where $\Psi_i$ are the shapes, while the $\Phi_i$ are polynomials in the
primary invariants. These polynomials are trivially extended from the 1D case
along the lines of Eq.~\eqref{triv}. Explicitly, let $t_m$ denote one-node
single-particle wave functions along the $x$-axis, e.g. $t_m=e^{2i\pi x_m/L}$
with $0<x<L$ for $N$ particles in a box, $m=1,\ldots,N$. Then all the primary
invariants in the $x$-direction are generated by Euler's
expression~\eqref{eulergen}, in other words they are all unrestricted
polynomials in the elementary symmetric functions $e_k$~\eqref{ek}. Replacing
$t$ by $u$ for the $y$-direction, and so on, the independent primary
invariants are just all the monomials generated by multiplying together $d$
copies of Eq.~\eqref{eulergen}, one for each direction in space. The functions
$\Phi_i$ above are arbitrary polynomials in these monomials, with complex
coefficients. [It is most convenient, although not necessary, to interpret
powers of the $e_k$ in these monomials without cross terms, e.g.
$e_1^2(t_1,t_2)\to t_1^2+t_2^2$.]

In this context the $e_k$ are called \emph{Euler bosons}. They are the same
for bosons and for fermions. Thus the shapes $\Psi_i$ alone encode both the
space dimension and the distinction between bosons and fermions. Euler bosons
have no zero-point energy, because it is contained in the shapes, which play
the role of vacua to Euler-boson excitations.

The insights above, based on the Hilbert series~\eqref{zdnq}, are restrictive
enough to obtain the wave functions of the shapes explicitly.\cite{Sunko16-1}
The construction is recursive and exhaustive. For each grade, one first
constructs all states based on shapes of lower grade, using the
formula~\eqref{genspan} with all possible $\Phi_i\neq 1$. The shapes are then
obtained as the orthogonal complement in the Hilbert space of that grade,
which is independently generated by Slater determinants\footnote{Slater
determinants are replaced by permanents (lose the alternating sign) for
identical bosons.} in the standard way. This algorithm makes it immediately
obvious that the dimension of the Hilbert-space algebra is indeed $N!^{d-1}$,
i.e.\ the number of shapes cannot be further reduced. Namely, no shape at a
given grade can be expressed in the form~\eqref{genspan} with $\Phi_i\neq 1$,
because all the states which can be so expressed are constructed explicitly
\emph{before} taking their orthogonal complement. Of course, the
orthogonal-complement procedure defines the shapes of a given grade only up to
linear basis transformations among themselves. That freedom is trivial
theoretically, but may be important in applications.

Unfortunately the above general algorithm is quite inefficient in practice,
because it requires the computation of the complete Hilbert spaces of a given
grade, which are much larger than the number of shapes. There is a much more
efficient algorithm, which is however only applicable to fermions in odd
dimensions.\cite{Sunko16-3}

\section{Discussion}

The developments reviewed above are quite recent, so that the present
discussion is necessarily more prospective than retrospective. There seems to
be more work than any one person or group can do to explore all possible
ramifications of the shape paradigm. Here only three which come to mind
immediately are briefly mentioned.

\subsection{Slater determinants in finite systems}

While Slater determinants still appear as an auxiliary tool in the algorithm
which constructs shapes, one feels that they have become redundant in physical
descriptions, at least of finite systems. There is a mathematical way to
express this feeling. Specialize to one dimension, where the only shape is the
ground-state Slater determinant $\Psi_0$. One may ask, what is $\Phi$ in the
expression $\Psi=\Phi\Psi_0$, if $\Psi$ is an excited-state Slater
determinant? We have seen above that the most convenient way to write $\Phi$
is in terms of Euler bosons, but then $\Psi$ is \emph{not} a Slater
determinant, in general. If it is, then the ratio $\Psi/\Psi_0$ is called a
Schur function, which is the generating function of non-standard Young
tableaux.\cite{Stanley99} The physical meaning of such an object is quite
obscure, and so is, by extension, the physical meaning of a generic Slater
determinant. By contrast, the elementary symmetric functions (Euler bosons)
have simple physical connotations in general. Thus $e_1$ in Eq.~\eqref{ek} is
a sum of one-body wave functions, which is physically a liquid, while $e_N$ is
their product, physically a gas.

While the lowest-graded shapes are always just the Slater determinants of the
(possibly degenerate) ground state, for $d>1$ shapes are generally
superpositions of Slater determinants with roughly equal amplitudes. With
complexity increasing by grade, they are the natural choices for strongly
correlated trial wave functions. Conversely, if the DFT approach gives
realistic one-body wave functions, the shape paradigm reduces to the
Slater-determinant one, at least for the ground state. Even in that case,
shapes may increase the scope of calculations of excited states and related
radiation transitions, which are an independent indicator of how realistic the
wave function is.

\subsection{Bands in spectra}

A spectrum is organized into bands when states of the same band are easily
connected by excitations, while states of different bands are not, i.e.\ they
behave as if they were of qualitatively different kinds. The lowest state in
each band is called a bandhead. There is a natural physical picture behind
this scheme: bandheads correspond to different configuratons of particles in
laboratory space. Some of these may be easy to excite by vibration, others by
rotation, while it is difficult to change a configuration from one prone to
vibration to another, prone to rotation.

The abstract structure of ideals of many-body Hilbert space, described in
Eq.~\eqref{ideal}, matches the above physical pattern. The band may be
understood mathematically as the ideal generated by the bandhead. The ideal is
closed under wave-function superposition, and excitations within it can be
described by multiplying the bandhead with Euler bosons. Physically, this
multiplication corresponds to bosonic excitations of the bandhead. In this way
one can distinguish theoretically between excitations within a band, which are
bosons, and excitations from one band to another, which involve
reconfigurations of the bandhead. The latter cannot be obtained
multiplicatively, because shapes are like prime numbers, $\Psi\neq\Phi\Psi'$
for any two distinct shapes $\Psi$ and $\Psi'$: one is not divisible by the
other.

In order for such a mathematical interpretation to have physical meaning, the
interaction matrix elements between the shapes should be small. This
proposition has been checked\cite{Sunko16-1} in small numerical examples in
$d=2$ and $d=3$ with the Coulomb interaction $V_C$, for a harmonic oscillator
potential (spin-polarized electrons in a quantum dot). It has been found that
$\left<\Psi_i\right|V_C\left|\Psi_j\right>\ll
\left<S_i\right|V_C\left|S_j\right>$, typically by about two orders of
magnitude, between any two shapes $\Psi_{i,j}$ and any two Slater determinants
$S_{i,j}$ of the same grade. This result is of course just a sanity check, as
the opposite outcome would have doomed the interpretation immediately.

The ideal structure explains why the original idea of Bloch\cite{Bloch34} to
represent excitations of fermionic systems as density waves was generically
successful only in one dimension.\cite{Tomonaga50} Bosonisation corresponds to
multiplying the ground state by a symmetric function. It can describe all
excitations in one dimension, where there is only one shape, the ground-state
Slater determinant, or all within one band for $d>1$. In general, however,
excitations may also involve a reconfiguration of the ground state, which is
an \emph{additive} correction as explained above, e.g.
$\Psi\to\Psi+\Phi\Psi'$, where $\Psi$ and $\Psi'$ are ``relatively prime.''
Thus the difficulty in representing excitations of fermionic systems by bosons
appears as just another case of the incompatibility between addition and
multiplication, ubiquitous throughout mathematics, of which, say, translation
and rotation are a particularly well-known example.

\subsection{The fermion sign problem}

The fermion sign problem\cite{Loh90} appears in simulations of fermionic
many-body systems in real space. It is most easily understood in a
sum-over-histories formulation. When a transition matrix element is formed as
a sum over trajectories in fermion many-body space, one has to distinguish
contributions which interfere destructively for physical reasons, from those
which should interfere constructively but have opposite signs because a phase
convention for the many-body wave function has not been implemented
consistently across many incremental updates of different
trajectories.\cite{Iglovikov15}

The sign problem appears only because there is more than one shape. In one
dimension, where there is only one shape $\Psi_0$, the factorization
$\Psi=\Phi\Psi_0$ immediately translates the fermionic problem into a bosonic
one involving the symmetric functions $\Phi$. Physically, in one dimension it
is possible to fix a phase convention simply by forbidding particles to skip
over each other in a simulation. If they never exchange places, the difference
between bosons and fermions cannot manifest itself either, so the orbital
solution is actually a symmetric function for both fermions and bosons, the
difference between the two cases being in the fixed factor $\Psi_0$.

The representation~\eqref{genspan} provides a generalization of this insight
to $d>1$. Instead of writing the wave function as in~\eqref{genspan}, one can
list the coefficients $\Phi_i$ as a vector-like object,
\be
\left(\Phi_1,\Phi_2,\ldots,\Phi_{N!^{d-1}}\right).
\label{fmod}
\ee
When the $\Phi_i$ are symmetric polynomials, such a structure is called a
\emph{free module} (as distinct from a vector space, where the $\Phi_i$ would
be just numbers). It carries the idea that the shapes are a basis to its
natural conclusion, proposing to calculate in the space~\eqref{fmod} of
``coordinates'' $\Phi_i$ instead of directly in the space~\eqref{genspan} of
wave functions $\Psi$. Calculations in the free module have no sign problem
\emph{a priori}, because the $\Phi_i$ are symmetric functions. It remains to
be seen whether they will be practical. At least, there is a chance to manage
complexity, because one can imagine\cite{Sunko16-1} truncating the
list~\eqref{fmod} to limit the calculation to one ideal, or band of
excitations. Such an approach is similar in spirit to a fixed-node
approximation\cite{Bressanini12,Filinov14} in usual simulations.

The proposal to use the free module in order to avoid the sign problem can be
placed in the more general setting of \emph{constraint
elimination}.\cite{Derksen02} In some design problems constraints can be
encoded as requirements that certain polynomials $g_i$ simultaneously evaluate
to zero, leading to the expression of allowed solutions in the
form~\eqref{palg}. Eliminating constraints means mapping the original space of
the problem to some reduced space in which they are satisfied automatically.
The whole geometric locus of zeros in the original space (\emph{kernel} of
this mapping) maps to a single point, zero, in the reduced space. This is the
case when wave functions are mapped to the free module: $\Psi=0$ in
Eq.~\eqref{genspan} corresponds to $(0,0,\ldots,0)$ in Eq.~\eqref{fmod}.

Hilbert's \emph{Nullstellensatz}\cite{Derksen02} (``zeroplacesclaim'') is an
either/or proposition about constraint elimination: either the constraints
$g_i=0$, $i=1,\ldots,D$, are incompatible, or every solution $f$ of the
problem can be written as an algebraic root of the expression~\eqref{palg},
$f^m=p$ with an integer $m\ge 1$. The many-body problem is an interesting
variation of this general framework. Normally the simultaneous solutions of a
set of polynomial equations are a discrete set of points, called an
\emph{algebraic variety}. Each shape vanishes according to the Pauli principle
when $\vec{r}_i=\vec{r}_j$ with $i\neq j$ for any two particles,
$i,j=1,\ldots,N$. These points of coincidence are the simultaneous zeros of
the shapes, corresponding to the algebraic variety of the problem. However, as
noticed long ago in the many-body context,\cite{Ceperley91} when $d>1$ the
algebraic variety is a subset of a larger object, the \emph{nodal surface} of
the total wave function $\Psi$. The nodal surface is a (possibly
multiply-connected) differentiable manifold defined by $\Psi=0$, on which each
individual shape need not vanish. It appears only because the Pauli principle
is triggered by an equality of \emph{vector} variables, a weaker condition for
zeros than explicitly envisaged by Hilbert. The nodal surface is the reason
for the practical difficulty in defining the wave-function phase consistently
in simulations, because it is not known in advance. Thus it is of considerable
interest to understand its relation to the algebraic variety. Of course, a
polynomial like~\eqref{palg} will in general have more zeros than there are
common zeros of the $g_i$. However, for $d=1$ the extra zeros are discrete and
separate from the algebraic variety. The specific feature of the problem for
$d>1$ is that the algebraic variety naturally generates a whole
differential-geometric ``zero object,'' the nodal surface, which contains it.

The free module~\eqref{fmod} is a structured implementation of Heisenberg's
matrix mechanics, which refines Dirac's vector-space interpretation.
Heisenberg's algebraic approach\cite{Heisenberg26} eschewed wave-function
\emph{arguments} --- real-space coordinates --- in favor of wave-function
\emph{values}, as encoded by the matrix elements. The textbook harmonic
oscillator is a one-body realization of this program, where replacing $x\pm
d/dx$ by the creation and destruction operators faithfully maps laboratory
space onto wave-function space, so that one never needs to revert to the
variable $x$ for concrete calculation. In close analogy, the $\Phi_i$ are
polynomials in the elementary symmetric functions~\eqref{ek}, which are
themselves expressed directly in terms of the \emph{values} of the one-body
wave functions. Changing powers of the $e_i$ in this formalism corresponds to
harmonic-oscillator ladder operations. The representation~\eqref{fmod} of the
free module thus opens the possibility of practically implementing
Heisenberg's approach in a general many-body setting. This program has a much
larger scope than the motivating fermion sign problem.

\section{Conclusions}

The general algebraic structure of many-body Hilbert space described in this
review provides a firm mathematical foundation to study the physics of finite
systems in a pleasantly intuitive language, with suggestive correspondences
between mathematical structures and observations. Physical vacua appear
mathematically as generators of ideals in the algebra of many-body invariants,
while these ideals physically correspond to bands in the excitation spectra.
Although the required mathematics far exceeds contemporary training of physics
students in abstract algebra, such is remarkably not the case with students of
engineering. Industrial design optimization uses engineering constraints as
generators of allowed solutions in much the same way as shapes are used
here.\cite{Lv12} Such a large body of practical knowledge gives rise to
optimism for the proposed physical applications.

At the same time, specifics of the physical many-body problem should not be
overlooked. In particular, the arrangement of $Nd$ variables in $N$ columns
(vectors) of $d$ rows, however natural when describing particles, does not
seem to have been pursued in the mathematical literature on invariants.
Monographs\cite{Derksen02} and textbooks\cite{Stanley99} typically treat
all variables on an equal footing, corresponding to the many-body problem in
one dimension. Due to this additional structure in the variables when $d>1$,
the shape invariants described here appear to be new to mathematics as well,
although the author cannot claim so with authority.

Perhaps the most interesting insight based on the shape paradigm at present is
the reinterpretation of matrix mechanics in terms of a free module rather than
a vector space. It is tantalizing to speculate how far the young Heisenberg
would have pursued his algebraic approach if Schr\"odinger's analytic work had
never appeared, or if Hilbert or Noether had been the one to search for their
common ground. Hopefully the present brief review will encourage readers to
explore shapes from the point of view of their own expertise. 

\section*{Acknowledgements}

I thank D.~Svrtan for his help and interest. This work was supported by the
Croatian Ministry of Science grant 119-1191458-0512 and by the University of
Zagreb grant 202759.

\end{document}